\documentstyle[prl,aps,multicol,epsfig,amsmath]{revtex}
\begin{document}

\draft
\date{July 2001}
\title {Manipulation of  photon statistics of highly degenerate chaotic radiation
 }
\author{M. Kindermann$^a$, Yu.\ V. Nazarov$^b$, and C.\ W.\ J.\ Beenakker$^a$}

\address{$^a$Instituut-Lorentz, Universiteit Leiden, P.O. Box 9506, 2300 RA
Leiden, The Netherlands}

\address{$^b$Department of Applied Physics and Delft Institute of
Microelectronics and Submicrontechnology,\\ Delft University of
Technology, Lorentzweg 1, 2628 CJ Delft, The Netherlands}

\widetext

\maketitle

\begin{abstract}
Highly degenerate chaotic radiation has  a Gaussian density matrix and a large  occupation number of modes $f $. If it is passed through a weakly transmitting barrier, its counting statistics is close to Poissonian. We show that a second identical barrier, in series with the first, drastically modifies the statistics. The variance of the photocount is increased above the mean  by a factor $f$ times a numerical coefficient.  The photocount distribution   reaches a  limiting form with a Gaussian body and highly asymmetric tails. These are general consequences of the combination of weak transmission and multiple scattering. 
\end{abstract} 


\begin{multicols}{2}

\narrowtext

Chaotic radiation is the name given in quantum optics to a gas of photons that has a Gaussian density matrix \cite{Mandel}. The radiation emitted by a black body is a familiar example. The statistics of black-body radiation, as measured by a photodetector, is very close to the Poisson statistics of a gas of classical independent particles. Deviations due to photon bunching exist and were studied long ago by Einstein \cite{Einstein}. These are, however,  small corrections. To see effects of Bose statistics one needs a { \em degenerate} photon gas, with an  occupation number $f$  of the modes that is $ \gtrsim 1 $. Black-body radiation at optical frequencies is non-degenerate to a large degree ($f \simeq e^{-\hbar \omega/ k T} \ll 1$), even at temperatures reached on  the surface of the Sun. 

The degeneracy is no longer restricted by frequency and temperature if the photon gas is brought out of thermal equilibrium. The coherent radiation from a laser would be an extreme example of high degeneracy, but the counting statistics is still Poissonian because of the  special properties of a coherent state \cite{Mandel}.
 One way to create non-equilibrium chaotic radiation is spectral filtering  within the quantum-limited line width  of a   laser \cite{Exter}. This will typically be single-mode radiation. For multi-mode radiation one can pass black-body radiation through a linear amplifier. The amplification might be due to stimulated emission by an inverted atomic population or to stimulated Raman scattering \cite{Henry}. Alternatively, one can use the spontaneous emission from an amplifying medium that is well below the laser threshold  \cite{Beenakker1}. 

\begin{figure}
\centering\epsfig{file=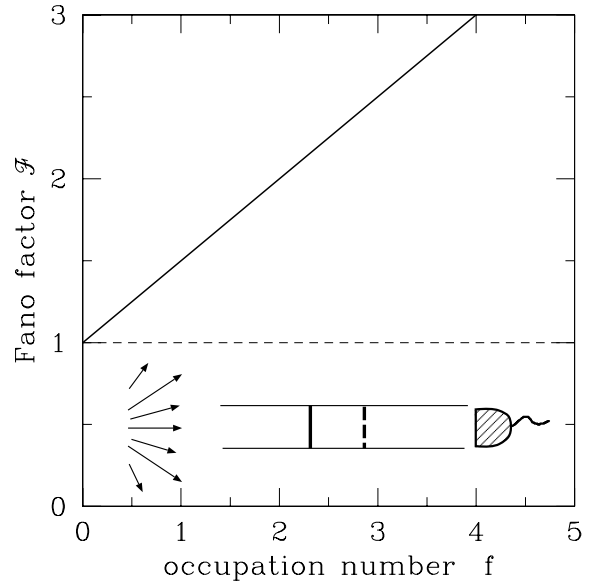,clip=,width=\linewidth}
\caption{Dependence of the Fano factor  on the  occupation number of the modes, for transmission through one (dashed line) or two  barriers (solid line). The inset shows schematically the photodetector (shaded) and the waveguide containing one or two barriers. }
\label{figure1}
\end{figure}

The purpose of this paper is to show that the statistics of  degenerate chaotic radiation can be manipulated by introducing scatterers, to an extent that would be impossible for both non-degenerate chaotic radiation and degenerate coherent radiation.  We will illustrate the difference by examining in some detail a simple geometry consisting of one or two weakly transmitting barriers embedded in a waveguide (see Fig.\ \ref{figure1}). For the single barrier the photocount distribution is close to Poissonian.  The mean photocount $\bar{n}$ is only changed by a factor of two upon insertion of the second barrier. But the fluctuations around the mean are greatly enhanced, as a result of multiple scattering in a region of large occupation number. We find that the distribution $P(n)$ for the double-barrier geometry is not only much broader than a Poisson distribution, it also has a markedly different shape.  
 
We consider a source of chaotic radiation that is not in thermal equilibrium. Chaotic radiation is characterized by a Gaussian density matrix $\rho$ in the coherent state representation \cite{Mandel}. For a single mode it takes the form 
\begin{equation} \label{eq:dens}
  \rho = \int{d \alpha d\alpha^{*}    (\pi \mu)^{-1} \exp (- \alpha^* \mu^{-1} \alpha)    |\alpha\rangle\langle\alpha|},
\end{equation}
where $\mu$ is a positive real number and $|\alpha\rangle$ is a coherent state (eigenstate of the photon annihilation operator $a$) with complex eigenvalue $\alpha$. If one takes into account more modes, ${ \alpha}$ becomes a vector ${\boldsymbol  \alpha}$ and ${ \mu}$ a matrix ${\boldsymbol  \mu}$ in the space of modes. (The factor $  \pi \mu$ then becomes the determinant $|| \pi {\boldsymbol  \mu}  ||$.)
We take a waveguide geometry  and assume that the radiation is restricted to a narrow frequency interval $\delta \omega$ around $\omega_0$. In this case the indices $n$,$m$ of $\alpha_n$, $\mu_{mn}$ label the $N$ propagating waveguide modes at frequency $\omega_0$. 

In thermal equilibrium at temperature $T$, the covariance matrix ${\boldsymbol  \mu}= f \openone $ equals the unit matrix $\openone$ times the scalar factor $f=(e^{\hbar \omega/k T}-1)^{-1}$, being the Bose-Einstein distribution function. Multi-mode chaotic radiation out of thermal equilibrium has in general a non-scalar ${\boldsymbol  \mu}$. 
We assume that $\boldsymbol \mu$ is a property of the amplifying medium, independent of the scattering properties of the waveguide to which it is coupled. Feedback from the waveguide into the amplifier is therefore neglected.

The radiation is fully  absorbed at the  other end of the waveguide  by a photodetector.  We seek the probability distribution $P(n)$ to count $n$ photons in a time $t$. It is convenient to work with the cumulant generating function $ F(\xi) = \ln[\sum_n{e^{\xi    n} P(n)}]$. For long counting times $t \delta \omega \gg 1$ it is given by the Glauber formula \cite{Mandel,Glauber}
\begin{equation} \label{eq:Glauber}
F(\xi) = \frac{t \delta \omega}{2 \pi} \ln   {\rm  Tr}\;  \bigl( \rho   : \exp[(e^{\xi}-1)  {\bf a}^{\dag}_{\rm out} {\bf a}_{\rm out}]: \bigr).
\end{equation}
Here ${\boldsymbol a}_{{\rm out}} $ is the vector of annihilation operators for the modes going out of the waveguide and into the photodetector. The colons $: \;\;   :$ indicate normal ordering (creation operators to the left of annihilation operators). The transmission matrix ${\boldsymbol t}$ relates ${\boldsymbol a}_{\rm out} = {\boldsymbol{ t a}}$ to the vector ${\boldsymbol a}$ of annihilation operators entering the waveguide. Substituting Eq.\ (\ref{eq:dens}) for $\rho$, we find 
\begin{eqnarray} \label{eq:statistics}
 F(\xi)&  = & \frac{t  \delta \omega}{2 \pi}   \ln   \int{ d {\boldsymbol \alpha}   d {\boldsymbol \alpha}^*    ||\pi {\boldsymbol \mu}  ||^{-1}   \exp(- {\boldsymbol{\alpha}^{*}\boldsymbol{ \mu}^{-1} \boldsymbol{\alpha}})} \nonumber \\
       &    & \mbox{}  \times   \exp[ (e^{\xi} -1) {\boldsymbol{ \alpha}^{*} \boldsymbol{t}^{\dag} \boldsymbol{t \alpha}} ] \nonumber \\
       & =  & - \frac{t \delta \omega}{2 \pi}     \ln  || \openone - (e^{\xi}-1)   {\boldsymbol{ \mu   t}^{\dag}\boldsymbol{ t}} ||.
\end{eqnarray}

In thermal equilibrium, when ${\boldsymbol \mu} = f \openone$, the determinant can be evaluated in terms of the eigenvalues $T_n$ of the matrix product ${\boldsymbol{ t}^{\dag} \boldsymbol{t}}$. The resulting expression \cite{Beenakker1,Beenakker3}
\begin{equation} \label{eq:cumu}
 F(\xi) = -  \frac{t \delta \omega}{2 \pi} \sum_{n=1}^N   \ln[1 - (e^{\xi} -1)   f  T_n)  ]
\end{equation}
has a similar form as the generating function of the electronic charge counting distribution at zero temperature \cite{Lesovik,Levitov}, 
\begin{equation} \label{eq:el}
F_{\rm electron} (\xi) = \frac{teV}{2 \pi \hbar} \sum_{n=1}^N \ln[1 + (e^{\xi} -1) T_n],
\end{equation}
where $V$ is the applied voltage. 

The similarities become even more striking if one derives Eqs.\ (\ref{eq:cumu}) and (\ref{eq:el}) using the Keldysh approach \cite{Nazarov,Belzig}, which allows to analyse more general situations. In this approach, the cumulant generating function
\begin{eqnarray}  \label{eq:genf}
 F(i \xi)& =& \pm \frac{t}{4 \pi} \sum_n{ \int{d \omega \;\;{\rm Tr} \; \ln \bigl[ 1  }}  \nonumber \\
&& \mbox{} + \case{1}{4} T_n (\{ \check{G}_{\rm l} ,  e^{-i \xi \sigma_3  /2} \check{G}_{\rm r}  e^{i \xi \sigma_3  /2}\} +2) \bigr] 
\end{eqnarray}
is expressed in terms of the anticommutator of the   Keldysh matrix Green's functions $\check{G}_{\rm l,r}$ \cite{Rammer}. (The subscripts l,r refer to particles to the left or to the right of the scattering region.)  The matrix $\sigma_3$ is the third Pauli matrix in Keldysh space. The overall sign is $-$ for bosons and $+$ for fermions.

If the eigenvalues of ${\boldsymbol{t \mu t}^{\dag} }$ are $ \ll 1$ , we may expand the logarithm in Eq.\ (\ref{eq:statistics}) to obtain $F(\xi) = \bar{n}(e^{\xi}-1)$, with mean photocount
$\bar{n}= (t \delta \omega /2 \pi)  {\rm Tr}\; {\boldsymbol{ \mu  t}^{\dag}\boldsymbol{t} }$. The corresponding photocount distribution is Poissonian,
\begin{equation} \label{eq:Poisson}
P_{\rm Poisson}(n) = \frac{1}{n!} \bar{n}^n e^{-\bar{n}}.
\end{equation}
 In thermal equilibrium the deviations from a Poisson distribution will be very small, because the Bose-Einstein function is $ \ll 1$ at optical frequencies for any realistic  temperature. There is no such restriction on the covariance matrix  $\boldsymbol \mu$ out of equilibrium. This leads to striking deviations from Poisson statistics.

As a measure for deviations from a Poisson distribution we  consider the deviations from unity of the Fano factor. From Eq.\ (\ref{eq:cumu}) we derive
\begin{equation} \label{eq:Fano}
{\cal F} = \frac{{\rm Var}\; n}{\bar{n}} = 1 + \frac{{\rm Tr} \;({\boldsymbol{ \mu t}^{\dag}\boldsymbol{ t}})^2}{{\rm Tr}\; {\boldsymbol{ \mu t}^{\dag}\boldsymbol{ t}}}.
\end{equation}
A Fano factor ${\cal F} >1$ indicates  photon bunching. For example, for black-body radiation ${\cal F} = 1 + f$ \cite{Einstein}.  One might surmise that photon bunching is negligible if the waveguide is weakly  transmitting, so that $N^{-1} {\rm Tr} \; {\boldsymbol{ t}^{\dag}\boldsymbol{ t}} \ll 1$. That is correct if the weak transmission is due to a single  barrier. Then each transmission eigenvalue $T_n \ll 1$, hence ${\cal F} \approx 1$. However, if a second identical barrier is placed in series with the first one a remarkable increase in the Fano factor occurs.

Let us first demonstrate this effect for a scalar ${\boldsymbol \mu}= f \openone$, when it has a well-known electronic analogue \cite{Chen,Blanter}. We assume that $N \gg 1$ so that we may replace traces in Eq.\ (\ref{eq:Fano}) by integrations over the transmission eigenvalue $T$ with density  $\rho(T)$, 
\begin{equation}
 {\cal F} = 1 + f  \frac{\int_0^1{dT \rho(T) T^2}}{\int_0^1{dT \rho(T) T}
}.
\end{equation}
For a single barrier $\rho(T)$ is sharply peaked at a transmittance $\Gamma \ll 1$. Hence, ${\cal F} \approx 1$ for a single barrier. 
   For two identical barriers in series the density is bimodal \cite{Melsen},
\begin{equation} \label{eq:density}
 \rho(T) = \frac{N \Gamma}{2 \pi}  T^{-3/2} (1-T)^{-1/2}, 
\end{equation}
with a peak near $T=0$ and at  $T=1$. 
From this distribution we find that 
\begin{equation}  \label{eq:refnoise}
 {\cal F} =  1 + \case{1}{2}  f .
\end{equation}
While  the second barrier reduces the mean photocount  by only a factor of two, independently of the  occupation number $f$ of the modes, it can  greatly increase  the Fano factor for large $f$ (see Fig.\ \ref{figure1}).  From the electronic analogue (\ref{eq:el}) we would find ${\cal F} = 1$ for a single barrier and ${\cal F} = 1 - \case{1}{2} = \case{1}{2}$  for a double barrier \cite{Chen}. We conclude that for electrons the effect of the second barrier on the mean current and the Fano factor are comparable (both being a factor of two), while for photons the effect on the Fano factor can be orders of magnitude greater than on the mean current for $f \gg 1$.

The result (\ref{eq:refnoise}) is ``universal'' in that changing the nature of the multiple scattering will only change the numerical coefficient $\case{1}{2}$. For example, multiple scattering by disorder would give ${\cal F} = 1 + \case{2}{3} f$, in analogy with the electronic result \cite{Buettiker,Nagaev}  ${\cal F} = 1 - \case{2}{3}  = \case{1}{3}$. The numerical coefficient will also change if we have a broad-band detector, when
\begin{equation}
  {\cal F} = 1 + \case{1}{2} \frac{\int{d\omega f^2(\omega)}}{\int{d\omega f(\omega)}}.
\end{equation}
For a Lorentzian frequency profile, with maximum $f_{\rm max}$, one has ${\cal F} = 1 + \case{1}{4} f_{\rm max}$.

We now generalize this result to a non-scalar $\boldsymbol \mu$. An extreme case is a covariance matrix of rank one having all eigenvalues  $\mu_n$ equal to zero except a single one. This would happen if the waveguide is far removed from the source, so that its cross-sectional area $A$ is smaller than the coherence area $A_c$ \cite{footnote}. Since ${\rm Tr}\;(\boldsymbol{ \mu t}^{\dag}\boldsymbol{ t})^2 = ({\rm Tr}\;\boldsymbol{ \mu t}^{\dag}\boldsymbol{ t})^2$ if $\boldsymbol{\mu}$ is of rank one, the Fano factor reduces to ${\cal F} = 1 + {\rm Tr}\;  \boldsymbol{\mu t}^{\dag} \boldsymbol{t}$. The trace of $ \boldsymbol{ \mu t}^{\dag} \boldsymbol{t}$ is $ \ll 1$ for both  a single and  double barrier geometry, hence a second  barrier has no large effect on the noise if $A \lesssim A_c$.

 More generally, for a non-scalar $\boldsymbol \mu$ the Fano factor (\ref{eq:Fano})  depends not just on the eigenvalues $T_n$ of $\boldsymbol{ t}^{\dag}\boldsymbol{ t}$, but also on the eigenvectors. We write $\boldsymbol{t}^{\dag}\boldsymbol{ t} = \boldsymbol{ U}^{\dag} \boldsymbol{\tau U}$, with $\boldsymbol{U}$ the unitary matrix of eigenvectors. We assume strong intermode scattering by disorder inside the waveguide. The resulting $\boldsymbol{U}$ will then be uniformly distributed in the unitary group, independent of $\boldsymbol{\tau}$ \cite{report}. For $N \gg 1$ we can replace the traces in numerator and denominator in Eq.\ (\ref{eq:Fano}) by integrations over $\boldsymbol{U}$,

\begin{equation}
{\cal F} = 1 +   \frac{\int{d\boldsymbol{U} {\rm Tr} \;   (\boldsymbol{\mu  U}^{\dag}\boldsymbol{  \tau   U })^2}}{\int{d\boldsymbol{U} {\rm  Tr} \;   \boldsymbol{\mu  U}^{\dag} \boldsymbol{  \tau   U} }}.
\end{equation}
The integrations over $U$ can be carried out easily \cite{report}, with the result
\begin{equation}
{\cal F} =   1 + \langle \boldsymbol{\mu}\rangle \langle \boldsymbol{\tau}\rangle + \langle \boldsymbol{\mu}\rangle \frac{\langle\!\langle  \boldsymbol{\tau}^2\rangle\!\rangle}{\langle  \boldsymbol{\tau} \rangle} + \langle  \boldsymbol{\tau} \rangle \frac{\langle\!\langle  \boldsymbol{\mu}^2\rangle\!\rangle}{\langle  \boldsymbol{\mu}\rangle}.
\end{equation}
Here  $\langle \boldsymbol{\mu}^p  \rangle = N^{-1}  {\rm Tr} \;  {\boldsymbol \mu}^p$,  $\langle \boldsymbol{\tau}^p  \rangle = N^{-1}  {\rm Tr} \;  {\boldsymbol \tau}^p$ denote the spectral moments and  $\langle \!\langle \boldsymbol{\mu}^p  \rangle  \! \rangle$,  $\langle\! \langle \boldsymbol{\tau}^p  \rangle \! \rangle$ the corresponding cumulants. [For example,  $\langle \!\langle \boldsymbol{\tau}^2  \rangle \! \rangle =  \langle  \boldsymbol{\tau}^2    \rangle - \langle  \boldsymbol{\tau}    \rangle^2$.] 

Instead of Eq.\ (\ref{eq:refnoise}) we now have for the double barrier geometry a Fano factor 
\begin{equation}
{\cal F} =   1 + \case{1}{2} \langle \boldsymbol{\mu}\rangle ( 1 + \kappa), \;\; \kappa=\Gamma \frac{\langle\!\langle  \boldsymbol{\mu}^2\rangle\!\rangle}{\langle  \boldsymbol{\mu}\rangle^2}.
\end{equation}
 We may estimate the magnitude of the correction $\kappa$ by noting that, typically, only $N_c \simeq A/A_c $ eigenvalues of $\boldsymbol \mu$ will  be significantly different from $0$. If we ignore the spread among these $N_c$ eigenvalues, we have $\langle  \boldsymbol{\mu}^2\rangle\approx (N/N_c) \langle  \boldsymbol{\mu}\rangle^2$, hence $\kappa \approx \Gamma(N/N_c -1)$. This correction will be negligibly small for $\Gamma \ll 1$, unless $\Gamma N \gtrsim N_c$. We note that the bimodal distribution (\ref{eq:density}), on which the universal result (\ref{eq:refnoise}) is based, requires the large-$N$ regime $N \gg 1/\Gamma$. For a non-scalar $\boldsymbol \mu$ the requirement for a universal Fano factor is more stringent: $1 \ll \Gamma N \ll N_c$.

In the final part of this paper we consider the full photocount probability distribution   $P(n) = (2 \pi)^{-1} \int_0^{2 \pi}{d \xi \exp[F(i \xi)-i n \xi] }$. For large detection time this integral can be done in saddle point approximation.  The result has the form $P(n)=\exp\bigl[ \bar{n} g(n/\bar{n}) \bigr]$. For small relative deviations of $n$ from $\bar{n}$ the function $g(n/\bar{n})$ can be expanded  to second order in $n/\bar{n}$.  Thus  the body of the distribution tends to a Gaussian for $t \rightarrow \infty$, in accordance with the central limit theorem.
 The same  holds for the Poisson distribution  (\ref{eq:Poisson}). However, the tails of $P(n)$ for degenerate radiation remain non-Gaussian and different from the tails of $P_{\rm Poisson}(n)$.

Let us first investigate this for a scalar $\boldsymbol{\mu}=f \openone$. Replacing the sum over $n$ in Eq.\ (\ref{eq:cumu}) by the integral $ \int_0^1 dT \rho(T)$, which is allowed in the large-$N$ limit, we find, using Eq.\ (\ref{eq:density}), the generating function 
\begin{equation} \label{eq:gendouble}
F(\xi) = \frac{t \delta \omega}{2 \pi} N \Gamma \bigl[1-\sqrt{1-(e^{\xi} - 1)f} \bigr].
\end{equation}
The corresponding $P(n)$ is the K-distribution that has appeared before in a variety of contexts \cite{Glauber,Beenakker3,Jakeman}. It is usually considered only for $f \ll 1$, as is appropriate for thermal equilibrium. In the regime $1 \ll f \ll \bar{n}$ of interest here it has the form
\begin{equation} \label{eq:K}
 P(n) = C \; \exp \left( - \frac{n}{f} - \frac{\bar{n}^2}{n f }\right),
\end{equation}
with a normalization constant $C = (\pi f \bar{n}  )^{-1/2} \exp(2 \bar{n} /f)$.  The essential singularity at $n=0$ is cut off below $\bar{n}/\sqrt{f}$, where the distribution saturates at $P(0)= \exp(-2 \bar{n} / \sqrt{f})$. 
One can easily check that the body of the distribution (\ref{eq:K}) reduces to a Gaussian  with variance $ \case{1}{2}f \bar{n}$. 
\begin{figure}
\centering\epsfig{file=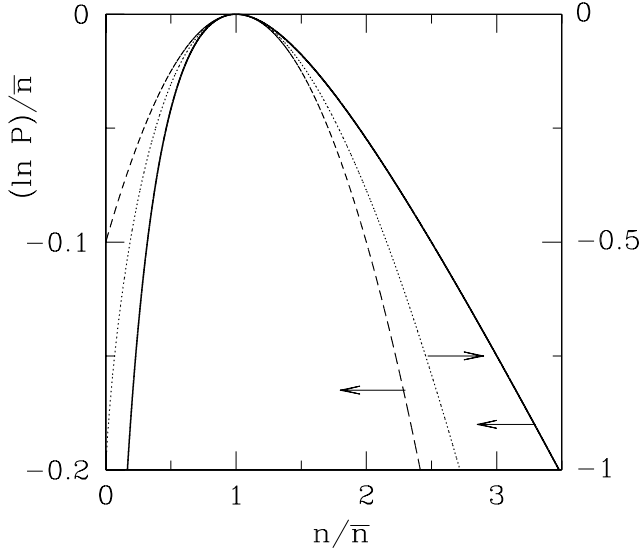,clip=,width=\linewidth}
\caption{Logarithmic plot of the photocount distribution for $f=8$ and $\bar{n} \to \infty$. The solid curve follows from Eq.\ (\ref{eq:gendouble})  (describing the double barrier geometry) and is very close to  the large-$f$ limit (\ref{eq:K}). The dashed curve is a Gaussian with variance $(1+ \case{1}{2} f) \bar{n}$  and the dotted curve is the Poisson distribution (\ref{eq:Poisson}). (Notice the different vertical scale for the dotted curve, chosen such that the Gaussian body of the Poisson distribution becomes evident.)   }
\label{figure2}
\end{figure}
In Fig.\ \ref{figure2} we compare the distribution (\ref{eq:K}) with a Gaussian and with a Poisson distribution, which has the asymptotic form $P_{\rm Poisson} = (2 \pi n)^{-1/2} \exp \bigl[ n - \bar{n} - n \ln(  n/\bar{n})   \bigr] $.   The logarithmic plot emphasizes the tails, which are markedly different. 

For a non-scalar $\boldsymbol{\mu}$ we find that the functional form of the large-$n$ tail depends only on the largest eigenvalue $\lambda_{\rm max} \gg 1$ of the Hermitian positive definite  matrix $ \boldsymbol{t \mu t}^{\dagger} $,
\begin{equation} \label{eq:tail}
 \lim_{n \to \infty} P(n) \propto e^{- n/\lambda_{\rm max}}.
\end{equation}
The number $\lambda_{\rm max}$ plays the role for a non-scalar $\boldsymbol{\mu}$ of the  filling factor $f$ in the result (\ref{eq:K}) for a scalar $\boldsymbol{\mu}$. (For broad-band detection  $\lambda_{\rm max}$ should also be maximized over frequency.) While the large-$n$ tail is exponential under very general conditions, the tail for $n \ll \bar{n}$ has no universal form.

In conclusion, we have calculated the effect of multiple scattering on the photodetection statistics of radiation that is both chaotic (like thermal radiation from a black body) and highly non-degenerate (like coherent radiation from a laser). Even for weak transmission there appear large deviations of the photocount distribution from  Poisson statistics, that are absent in the radiation from a black body or a laser. They take the form of an enhancement of ${\rm Var} \; n$ above  $\bar{n}$ by a factor $ \propto f$ and a slowing down of the large-$n$ decay rate of $P(n)$ by a factor  $1/f$. Explicit results have been given for a double barrier geometry, but these findings are generic and would apply also, for example, to multiple scattering by disorder. Because of this generality we believe that experimental observation of our predictions would be both significant and feasible. 

This work was  supported by the Dutch Science Foundation NWO/FOM.

\end{multicols}
\end{document}